\documentstyle[12pt,epsf]{article}
\begin{document}


\title{\bf `Real' vs `Imaginary' Noise in Diffusion-Limited Reactions}
\author{Martin J. Howard $^{\dag}$ and Uwe C. T\"auber $^{\ddag}$}
\date{
{\small\noindent{\it $^{\dag}$ CATS, The Niels Bohr Institute,
        Blegdamsvej 17, 2100 Copenhagen \O, Denmark.}\\}
{\small\noindent{\it $^{\ddag}$ Department of Physics, Theoretical
        Physics, University of Oxford, 1 Keble Road, Oxford, OX1 3NP,
        United Kingdom,\\ 
        and Linacre College, St. Cross Road, Oxford, OX1 3JA, United
        Kingdom.} 
\\ \today \\} 
}\maketitle
\vspace{-.4in}

\begin{abstract}
Reaction-diffusion systems which include processes of the form 
$A + A \to A$ or $A + A \to \emptyset$ are characterised by the
appearance of `imaginary' multiplicative noise terms in an effective
Langevin-type description. 
However, if `real' as well as `imaginary' noise is present, then
competition between the two could potentially lead to novel behaviour. 
We thus investigate the asymptotic properties of the following two
`mixed noise' reaction-diffusion systems.
The first is a combination of the annihilation and scattering
processes $2A \to \emptyset$, $2A \to 2B$, $2B \to 2A$, and 
$2B \to \emptyset$. 
We demonstrate (to all orders in perturbation theory) that this system
belongs to the same universality class as the single species
annihilation reaction $2A \to \emptyset$.
Our second system consists of competing annihilation and fission
processes, $2A \to \emptyset$ and $2A \to (n+2) A$, a model which
exhibits a transition between active and absorbing phases.
However, this transition and the active phase are not accessible to
perturbative methods, as the field theory describing these reactions
is shown to be non-renormalisable. 
This corresponds to the fact that there is no stationary state in the
active phase, where the particle density diverges at finite times. 
We discuss the implications of our analysis for a recent study of
another active / absorbing transition in a system with multiplicative
noise.
\

\begin{center}
\end{center}
\end{abstract}


\newpage

\section{Introduction}

Recently the effects of fluctuations in reaction-diffusion systems
have attracted considerable attention (for reviews, see
Refs.~\cite{KK,Cardy}).
In sufficiently low spatial dimensions the presence of microscopic
particle density fluctuations causes traditional approaches, such as
mean-field rate equations, to break down. 
This has led to the introduction of field-theoretic methods, based on
`Hamiltonian' representations of the associated classical master 
equation \cite{Doi,Grassberger,Peliti1}. 
These methods allow fluctuations to be handled in a systematic manner.
The first system to be analysed in this way was the single species
annihilation reaction $A + A \to \emptyset$, \cite{Peliti2,Lee}
where it was shown with renormalisation-group (RG) methods that for
dimensions $d < 2$, the average density $\overline {n(t)}$ decays to 
zero at large times according to the power law 
\begin{equation}
        \overline{n(t)} \sim E_d \, t^{-d/2} \ ,
\label{purann}
\end{equation}
with $E_d$ denoting a universal amplitude (for uncorrelated initial
conditions), while $\overline{n(t)} \sim E_2\,t^{-1} \ln t$ in $d=2$. 
Furthermore Peliti has demonstrated that the coagulation reaction 
$A + A \to A$ belongs to the same universality class as the pure
annihilation process $A + A \to \emptyset$ \cite{Peliti2}. 
Physically the anomalously slow decay of eq. (\ref{purann}) results
from the {\it anti}correlation of particles in low dimensions. 
Due to the `reentrancy' property of random walks for $d\leq 2$, once
two particles are in close proximity they will then tend to react
rather quickly. 
Hence at large times the remaining (unreacted) particles are likely to
be situated far from their nearest neighbours (i.e. the particles
become anticorrelated).

Markedly more complex behaviour may arise once particle production
processes are also permitted. 
For example, the `Branching and Annihilating Random Walk' (BARW)
system defined by the reactions $2A \to \emptyset$ and 
$A \to (m+1) A$, displays a dynamic phase transition between an
`active' ($\overline {n(t)} \to n_s > 0$ for $t \to \infty$) and an
`inactive and absorbing' state ($\overline {n(t)} \to 0$ for 
$t \to \infty$), with a remarkable difference between the cases of odd
and even number of offspring $m$ \cite{GraKrau,TakTre}. 
For odd $m$ and $d \leq 2$ the transition is basically characterised
by the critical exponents of directed percolation (DP) 
\cite{Sundermeyer,Sugar,Janssen,Grass1}, whereas for even $m$ and 
$d < d_c' \approx 4/3$ the phase transition is described by a new
universality class, with the density in the entire absorbing phase
decaying according to the power law in eq.~(\ref{purann})
\cite{GraKrau,TakTre,Tauber}. 

A powerful method for the analysis of such systems is provided by the 
RG improved perturbation expansion \cite{Lee,Howard,Rey}. 
However, once a field-theoretic action for the system has been derived
(from a microscopic master equation), it is also possible to write down
effective Langevin-type equations, where the form of the noise can
now be specified precisely, without any recourse to assumptions and
approximations \cite{Cardy}. 
The nature of the noise can look somewhat peculiar in this
representation, for example in the $A + A \to \emptyset$ reaction we
have the exact equation for the field $a(x,t)$: 
\begin{equation}
        \partial_t a(x,t) = D \, \nabla^2 a(x,t) 
              - 2 \lambda \, a(x,t)^2 + a(x,t) \, \eta(x,t) \ ,
\label{lang}
\end{equation}
where $D$ is the diffusion constant, $\lambda$ the reaction rate, and 
\begin{equation}
        \langle \eta(x,t) \rangle = 0 \ , \quad 
        \langle \eta(x,t) \eta(x',t') \rangle = 
                - 2 \lambda \, \delta^d(x-x') \delta(t-t') \ . 
\label{noise}
\end{equation}
Hence the noise $\eta$ is imaginary, a rather counterintuitive result. 
Recently Grinstein, Mu\~noz and Tu \cite{Grinstein} have studied an
equation superficially similar to that given above, with the aim to 
model active / absorbing transitions in autocatalytic chemical
processes. 
In the special case of a scalar field with a quadratic nonlinearity,
their model is defined by the equation
\begin{equation}
        \partial_t a(x,t) = D \, \nabla^2 a(x,t) - r \, a(x,t) 
                        - u \, a(x,t)^2 + a(x,t) \, \eta(x,t) \ ,
\label{GE}
\end{equation}
where
\begin{equation}
        \langle \eta(x,t) \rangle = 0 \ , \quad
        \langle \eta(x,t) \eta(x',t') \rangle = 2 \nu \,
                \delta^d(x-x') \delta(t-t') \ .
\label{GEnoise}
\end{equation}
Notice that the noise in eq.~(\ref{GEnoise}) has the opposite sign to
that considered previously (i.e. the noise of Ref.~\cite{Grinstein} is
{\it real}). 
However, this is an important point since a positive sign in the noise
correlator leads to divergences in the renormalised parameters of the
theory: 
\begin{equation}
\begin{array}{r}
        \nu_R=Z\nu \\ u_R=Zu
\end{array} \quad {\rm with} \quad Z = {1 \over 1 - \nu I_d(r)} \quad
        {\rm and} \quad I_d(r) = 
                \int{d^dk \over (2\pi)^d} {1 \over r + D k^2} \ .
\label{GErg}
\end{equation}
Hence new singularities emerge when the denominator of $Z$ vanishes. 
These divergences have not been present in earlier field-theoretic
studies of reaction-diffusion systems. 
Certainly if the noise in the model of Ref.~\cite{Grinstein} were
resulting from the reaction $A + A \to \emptyset$, its correlator
should have a {\it negative} sign, as described above (and hence 
{\it no} extra divergences would appear, rendering much of the
interesting behaviour in Ref.~\cite{Grinstein} obsolete). 
It is therefore not clear to us how the real multiplicative noise of
eqs.~(\ref{GE}),(\ref{GEnoise}) could be the {\it only} type generated 
--- we believe that internal, imaginary reaction noise should 
generically be present as well. 
Consequently the physical mechanism behind the noise analysed in
Ref.~\cite{Grinstein} remains somewhat obscure. 

One of the objects of this letter is to see if equations similar to 
that analysed by Grinstein, Mu\~noz and Tu (with real noise, and hence
potentially novel behaviour) can be derived consistently for certain 
reaction-diffusion systems using field-theoretic methods. 
Potentially at least, in an emerging competition between `real' and
`imaginary' noise contributions, the `real' component might prevail in
certain circumstances, conceivably leading to the scenario discussed
in Ref.~\cite{Grinstein}.
However, our main finding here is that although we have analysed two
systems where the noise correlator has both positive and negative
components, we have been unable to recover the new features discussed
in Ref.~\cite{Grinstein}. 
In fact, in the first of our model systems, a combination of
two-particle annihilation and scattering processes for the species $A$
and $B$, the (`imaginary') reaction noise dominates the long-time
behaviour, which is described by the asymptotic power law
(\ref{purann}). 
Neither can the `imaginary' noise terms be neglected in our second
reaction-diffusion system, namely combined annihilation and fission
processes of a single species $A$.
In this system a perturbative RG analysis breaks down in the active
phase and at the dynamical phase transition separating it from the
inactive state.
Therefore although we can only address the inactive phase, which is
again governed by the pure annihilation model, we believe this system
cannot reproduce any of the features in Ref.~\cite{Grinstein}.

In the following Sec.~2, we present a more thorough discussion of how
`imaginary' noise terms emerge in processes dominated by two-particle
annihilation reactions. 
On the other hand, problems belonging to the directed-percolation (DP)
universality class, as described by Reggeon field theory
\cite{Sundermeyer,Sugar,Janssen,Grass1}, can be faithfully represented
by a simple Langevin equation for the local particle density with
`real' noise.
In Sec.~3, we next present our scattering / annihilation model, which
we first analyse to one-loop order, and then to all orders in
perturbation theory by solving the coupled Bethe-Salpeter equations
for the vertices. 
In Sec.~4, we proceed to discuss the annihilation / fission reaction
system, and show that while its properties in the inactive state may
be analysed using field-theoretic methods, this is not the case in the
active phase or at the dynamic transition itself.
Finally, we summarise and discuss our results in the light of the
recently proposed transition scenario for `real' multiplicative noise
problems \cite{Grinstein}.

\section{`Real' vs `imaginary' noise in reaction-diffusion systems}

Before turning to our investigation of models with competing `real'
and `imaginary' noise terms, we briefly outline and review the
general issue of how to systematically include fluctuation effects in
reaction-diffusion systems. 
Above the upper critical dimension, a qualitatively correct analysis
may be obtained from the associated mean-field rate equations for the
average particle densities.
Below this dimension, where fluctuations become important, it is 
tempting to apply a Langevin equation approach, motivated by the
success of this technique in equilibrium critical dynamics.
However, one has to be aware that these typically irreversible
reactions constitute a dynamical system far away from thermal
equilibrium. 
Thus there is no fluctuation-dissipation theorem available which could
serve as a guide to the appropriate form of the Langevin noise
correlations. 
One could of course just try the simplest ansatz, namely some form
of white noise multiplicatively coupled to a certain power of the
particle densities, in order to ensure that all fluctuations vanish
when there are no particles left (i.e., in the absorbing state).
But, as we shall see shortly, at least for processes dominated by
two-particle annihilation reactions, this generically leads to an 
incorrect analysis.

Thus, in order to systematically include the effects of microscopic
density fluctuations in low dimensions, one can instead start with the
corresponding classical master equation, then represent this stochastic 
process by the action of second-quantised bosonic operators, and
finally use a coherent-state path integral representation to map this
system onto a field theory. 
This mapping itself is a standard procedure, and is described in
detail in Refs.~\cite{Cardy,Grassberger,Peliti1,Lee}, for instance.
Apart from the continuum limit that is usually taken, this procedure
provides an {\it exact} mapping of the initial master equation, and
involves no assumptions whatsoever regarding the form of the noise, the
relevance or irrelevance of certain terms etc.
Note that the resulting bosonic theory applies only to systems where
there is {\it no restriction} on the particle occupation number in the
microscopic model. 
For the description of exclusion processes where the site occupation
numbers are restricted to $0$ or $1$, obviously a fermionic
representation is more useful.

For example, for the simple two-particle annihilation reaction 
$A + A \to \emptyset$ the ensuing field-theoretic action reads
\begin{equation}
	S = \int \! d^dx \int \! dt 
	\left[ {\hat a} (\partial_t - D \nabla^2) a 
		- \lambda (1 - {\hat a}^2) a^2 \right] \ ,
\label{annr1}
\end{equation}
where we omit boundary terms relating to the initial conditions and
the projection state.
Here $D$ denotes the diffusion constant, $\lambda$ the annihilation
rate, and ${\hat a}(x,t)$ and $a(x,t)$ are bosonic fields. 
The stationarity conditions (`classical field equations') 
$\delta S / \delta a=0$ and $\delta S / \delta {\hat a}=0$ yield,
respectively, ${\hat a} = 1$ and the mean-field rate equation
$\partial_t a = D \nabla^2 a - 2 \lambda a^2$.
Thus, within the mean-field approximation we can identify $a(x,t)$
with the local `coarse-grained' particle density.
However, fluctuations in $a(x,t)$ may {\it not} be simply related to
density variations, as can be seen by performing the shift 
${\hat a} = 1 + {\bar a}$,
\begin{equation}
	S = \int \! d^dx \int \! dt 
	\left[ {\bar a} (\partial_t - D \nabla^2) a 
	+ 2 \lambda {\bar a} a^2 + \lambda {\bar a}^2 a^2 \right] \ .
\label{annr2}
\end{equation}
Integrating out the `response' field ${\bar a}$ in the functional
integral $\int{\cal D}a{\cal D}\bar a \exp(-S)$ then leads precisely
to eq.~(\ref{lang}) with the {\it negative} noise correlator
(\ref{noise}).
Physically this counterintuitive result corresponds to the 
{\it anti}correlation of particles in low dimensions.  
Furthermore, power counting reveals that this noise, which originates
from the quartic term in the above action, becomes relevant below a
critical dimension $d_c=2$. 
Because of this pure imaginary noise, $a(x,t)$ clearly cannot
represent a physical density variable. 
Although $\langle a(x,t) \rangle$ {\it is} equal to the mean density
$\overline {n(x,t)}$, similar relations do not hold for higher
correlators, see Ref.~\cite{Cardy}. 
Moreover, since $a(x,t)$ is {\it not} the density field, this
means that the noise must also take a non-standard from --- in fact $\eta$
in eq.~(\ref{noise}) represents only the contribution to the overall
noise from the reaction process. 
In reality, of course, diffusive and reaction noise can never be
disentangled from one another, and consequently there is no particular
reason why the noise in eq.~(\ref{noise}) should be `real', i.e.,
described by a strictly positive correlator.

However, it is possible to give descriptions where the diffusive noise
does appear explicitly. 
One approach begins with a Langevin equation including both real
reaction noise {\it and} diffusive noise. 
In that case the field $a(x,t)$ represents a coarse-grained local
density. 
This is the approach taken, for example, by Janssen in
Ref.~\cite{Janssen}. 
A second possibility is to begin with the representation (\ref{annr1})
and then obtain an {\it equivalent} description in terms of `density'
variables by considering the canonical transformation
$a = \exp(-{\tilde \rho})\rho$, ${\hat a} = \exp({\tilde \rho})$,
${\hat a} a = \rho$ (see Ref.~\cite{Grass1}).
The resulting effective action in terms of the new `density' fields
$\rho$ and ${\tilde \rho}$ then includes a term 
$\propto - {\tilde \rho}^2 \rho^2$ which corresponds to pure real
noise.
However, the ensuing field theory contains an extra `diffusion noise'
contribution.

Hence we see that the `naive' Langevin equation (i.e.,
eq.~({\ref{lang}), but with a positive noise correlator and no
diffusion noise) does not provide an appropriate effective description
of the above system.
On the other hand, for the standard Gribov process $A \to \emptyset$,
$A \to A + A$, $A + A \to A$ \cite{Janssen,Grass1}, the action in
terms of the shifted fields reads
\begin{equation}
	S = \int \! d^dx \int \! dt 
	\left[ {\bar a} [\partial_t + D (r - \nabla^2)] a 
	- \sigma {\bar a}^2 a
	+ \lambda {\bar a} a^2 + \lambda {\bar a}^2 a^2 \right] \ .
\label{drper}
\end{equation}
Here $D$ and $\lambda$ represent the diffusion constant and
annihilation rate as before, $\sigma$ is the branching rate, and
$r = (\mu - \sigma)/D$ where $\mu$ denotes the spontaneous decay rate. 
Integrating out the response fields now yields
\begin{equation}
        \partial_t a(x,t) = D \, (\nabla^2 - r) a(x,t) 
	        - \lambda \, a(x,t)^2 + \eta(x,t) \ ,
\label{dplan}
\end{equation}
with
\begin{equation}
	\langle \eta(x,t) \rangle = 0 \, , \ 
	\langle \eta(x,t) \eta(x',t') \rangle = 2
		[\sigma \, a(x,t) - \lambda \, a(x,t)^2] \, 
		\delta^d(x-x') \delta(t-t') \ .
\label{dpnoi}
\end{equation}
It should be noted, however, that the effective coupling entering the
perturbation expansion is actually $\propto \sigma \lambda$, which
leads to an upper critical dimension $d_c = 4$. 
Consequently, the term $\lambda {\bar a}^2 a^2$ becomes irrelevant,
and we are left with a pure positive definite noise correlator 
$\propto \sigma a(x,t)$ (or `square-root' multiplicative noise, if we
replace $\eta$ with $\sqrt{a} \eta'$).
After a simple rescaling, the action is readily mapped onto Reggeon
field theory for directed percolation
\cite{Sundermeyer,Sugar,Janssen,Grass1}.
Thus, here we encounter the generic case where $a(x,t)$ {\it can} be
identified with a coarse-grained density field.
Physically, the above reactions lead to particle clustering, and local
densities indeed constitute a natural choice for the order parameter
field.
We also remark that the signs and magnitudes of the prefactors (which
may even be chosen to be imaginary) of the cubic nonlinearities in the
action (\ref{drper}) do not matter {\it provided} their product
$- \sigma \lambda$ remains real and negative.

In the above analysis, we have brought out the contrasting features of
two categories of reaction-diffusion systems --- containing either
imaginary (anticorrelating) or real (clustering) noise. 
However, reaction-diffusion systems with a multiplicative noise term
$\propto a\eta$ are characteristically governed by pair reaction
processes and thus have {\it `imaginary'} noise. 
This is in contrast to the ansatz in Ref.~\cite{Grinstein}.
Of course, one might argue that if there are {\it both} `real' and
`imaginary' noise contributions present, then the `real' parts might
prevail and lead to the scenario discussed in \cite{Grinstein}.
This possibility motivates the following two case studies of combined
scattering / annihilation and annihilation / fission reactions to
which we now turn our attention.

\section{The scattering and annihilation process}

The first reaction-diffusion system we want to consider consists of
the four reaction processes
\begin{equation}
        A + A \to \emptyset \ , \quad 
        A + A \to B + B \ , \quad 
        B + B \to A + A \ , \quad {\rm and} \quad
        B + B \to \emptyset \ ,
\label{scatt}
\end{equation}
which occur at rates $\lambda_{AA}$, $\lambda_{AB}$, $\lambda_{BA}$, 
and $\lambda_{BB}$, respectively, and with diffusion constants $D_A$
and $D_B$ for the $A$ and $B$ particles. 
We choose uncorrelated initial conditions where the $A$ and $B$
particles are distributed randomly. 
Physically the above reaction scheme might occur if the $A$ particles
could undergo a scattering process turning into $B$ particles, and
vice versa, in addition to the presence of the annihilation reactions.
In order to systematically include the effects of microscopic density
fluctuations in low dimensions, we represent the corresponding master
equation by a coherent-state path integral (see Sec.~2).
In terms of the continuous fields $a$, ${\bar a}$, $b$, ${\bar b}$,
the diffusivities $D_i\neq 0$, continuum reaction rates $\{\lambda_{ij}\}$,
and the initial homogeneous densities $n_i$ (where $i,j = A,B$), 
the action reads (for $t \geq 0$):
\begin{eqnarray}
        & & \hspace{-.2in} S = \int \! d^dx \int \! dt \left[ 
        {\bar a} (\partial_t - D_A \nabla^2) a +
        {\bar b} (\partial_t - D_B \nabla^2) b +
        2 \lambda_{AA} {\bar a} a^2 + \lambda_{AA} {\bar a^2} a^2 +
        \right. \nonumber \\ & & \left. +
        2 \lambda_{BB} {\bar b} b^2 + \lambda_{BB} {\bar b}^2 b^2 +
        2 \lambda_{AB} {\bar a} a^2 + \lambda_{AB} {\bar a^2} a^2 - 
        2 \lambda_{AB} {\bar b} a^2 - \lambda_{AB} {\bar b^2} a^2 +
        \right. \nonumber \\ & & \left. \quad +
        2 \lambda_{BA} {\bar b} b^2 + \lambda_{BA} {\bar b^2} b^2 -
        2 \lambda_{BA} {\bar a} b^2 - \lambda_{BA} {\bar a^2} b^2 -
        n_A {\bar a} \delta(t) - n_B {\bar b} \, \delta(t) \right] \ .
\label{scfth}
\end{eqnarray}
If we now integrate out the response fields $\bar a$ and $\bar b$ from
the functional integral 
$\int {\cal D}a {\cal D}{\bar a} {\cal D}b {\cal D}{\bar b} \exp(-S)$,
we find that the above reaction-diffusion system can be described
{\it exactly} by a pair of Langevin-type equations
\begin{eqnarray}
        & & \hspace{-.2in} \partial_t a(x,t) = D_A \, \nabla^2 a(x,t) 
        - 2 (\lambda_{AA}+\lambda_{AB}) \, a(x,t)^2 + 2 \lambda_{BA} \, 
              b(x,t)^2 + \eta_A(x,t) \, , \ \nonumber \\
        & & \hspace{-.2in} \partial_t b(x,t) = D_B \, \nabla^2 b(x,t)
        - 2 (\lambda_{BB}+\lambda_{BA}) \, b(x,t)^2 + 2 \lambda_{AB} \, 
              a(x,t)^2 + \eta_B(x,t) \, , \
\label{scalan}
\end{eqnarray}
with noise correlations
\begin{eqnarray}
        & & \qquad \qquad \qquad \langle \eta_A(x,t) \rangle = 
        \langle \eta_B(x,t) \rangle = 0 \ , \\
        & & \langle \eta_A(x,t) \eta_A(x',t') \rangle = 
        [\lambda_{BA} \, b(x,t)^2 - (\lambda_{AA}+\lambda_{AB}) \,
        a(x,t)^2] \, \delta^d(x-x') \delta(t-t') \ , \nonumber \\
        & & \langle \eta_B(x,t) \eta_B(x',t') \rangle =
        [\lambda_{AB} \, a(x,t)^2 - (\lambda_{BB}+\lambda_{BA}) \,
        b(x,t)^2] \, \delta^d(x-x') \delta(t-t') \ . \nonumber 
\label{scnois}
\end{eqnarray}
Hence, as desired, we have constructed a system where in a
Langevin-type formalism we have terms of {\it both} signs present in 
the correlator.
Thus we can now attempt to answer the question of whether this
`competition' alters the structure of the theory in low dimensions
where fluctuations are of vital importance.

Power counting on the action (\ref{scfth}) reveals that all the
reaction rates $\{\lambda_{ij}\}$ have dimension $\sim \mu^{2-d}$,
where $\mu$ denotes a momentum scale. 
Hence we expect to find a critical dimension $d_c=2$, below which
fluctuations change the mean-field behaviour qualitatively and the
theory must be renormalised.
As in the pure annihilation model (\ref{annr2}), this renormalisation
is simple since the diagrammatic structure of the theory does not
permit any dressing of the propagators.
Hence the only renormalisation required is that for the reaction
rates. 
We now define $\tilde\lambda_{AA}=\lambda_{AA}+\lambda_{AB}$, 
$\tilde\lambda_{AB}=\lambda_{AB}$, $\tilde\lambda_{BA}=\lambda_{BA}$
and $\tilde\lambda_{BB}=\lambda_{BB}+\lambda_{BA}$.
The temporally extended vertex function for $\tilde\lambda_{AA}(k,s)$
to one-loop order is given by the sum of diagrams shown in Fig.~1
(here $s$ is the Laplace transformed time variable; time runs from
right to left).
The diagrams for the other vertex functions look quite similar. 
Evaluation of these one-loop diagrams yields the following form of the
renormalised reaction rates: $g_{ij} = C_d 
\tilde\lambda_{ij}(k,s)|_{k^2/4=\mu^2,s=0} / D_i \mu^{\epsilon}$,
where $\epsilon = 2-d$, and $C_d = \Gamma(2-d/2) / 2^{d-1} \pi^{d/2}$
is a geometric factor. 
This leads in a straightforward manner to the following one-loop RG
beta functions $\beta_{ij} = \mu \, \partial g_{ij} / \partial \mu$: 
\begin{eqnarray}
        & & \beta_{AA} = g_{AA} 
                (- \epsilon + g_{AA}) + g_{AB} \, g_{BA} \ ,
\label{betaa} \\
        & & \beta_{AB} = g_{AB}
                (- \epsilon + g_{AA} + g_{BB}) \ ,
\label{betab} \\
        & & \beta_{BA} = g_{BA}
                (- \epsilon + g_{AA} + g_{BB}) \ ,
\label{betba} \\
        & & \beta_{BB} = g_{BB}
                (- \epsilon + g_{BB}) + g_{AB} \, g_{BA} \ .
\label{betbb}
\end{eqnarray}

\begin{figure}
\begin{center}
\leavevmode
\vbox{
\epsfxsize=4.5in
\epsffile{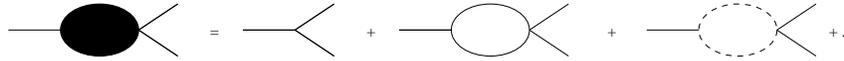}}
\end{center}
\caption{The temporally extended vertex function $\tilde\lambda_{AA}(k,s)$
        to one-loop order.} 
\end{figure}

In fact it can be shown that the above one-loop beta functions are
actually {\it exact} to all orders in perturbation theory. 
This is readily accomplished by writing down the full coupled
Bethe-Salpeter equations for the vertices.
Diagrammatically, this corresponds to replacing {\it either} the
right-hand {\it or} the left-hand bare vertices in all the one-loop 
contributions (see Fig.~1) by their fully renormalised counterparts.
This freedom of choice immediately implies the relation
\begin{equation}
	\tilde\lambda_{AB}(k,s) / \tilde\lambda_{AB} 
	= \tilde\lambda_{BA}(k,s) / \tilde\lambda_{BA} = N(k,s) \ .
\label{BS1} 
\end{equation}
After absorbing the diffusivities $D_i$ into the bare couplings
$\tilde\lambda_{ij}$ and the full vertex functions 
$\tilde\lambda_{ij}(k,s)$, respectively, and introducing the
abbreviation 
$I_i(k,s) = (2\pi)^{-d} \int d^dp~ [p^2 + (k^2/4) + (s/2D_i)]^{-1}$,
the coupled {\it exact} Bethe-Salpeter equations can be explicitly
written as 
\begin{eqnarray}
	& & \tilde\lambda_{AA}(k,s) \, [1 + \tilde\lambda_{AA} 
            I_A(k,s)] + \tilde\lambda_{AB}(k,s) \, \tilde\lambda_{BA}
            	\, I_B(k,s) = \tilde\lambda_{AA} \ ,
\label{BS2} \\
	& & \tilde\lambda_{AA}(k,s) \, \tilde\lambda_{AB} \, I_A(k,s) 
        + \tilde\lambda_{AB}(k,s) \, [1 + \tilde\lambda_{BB} 
		\, I_B(k,s)] = \tilde\lambda_{AB} \ ,
\label{BS3}
\end{eqnarray}
together with a second pair of equations which follow by interchanging
$A \leftrightarrow B$ in eqs.~(\ref{BS2}) and (\ref{BS3}).
These coupled linear equations (\ref{BS1})--(\ref{BS3}) for
$\tilde\lambda_{ij}(k,s)$ are solved by 
\begin{equation}
	N(k,s)^{-1} = [1 + \tilde\lambda_{AA} I_A(k,s)] \, 
	[1 + \tilde \lambda_{BB} \, I_B(k,s)] - 
	\tilde\lambda_{AB}\, \tilde\lambda_{BA}\, I_A(k,s)\, I_B(k,s)\ ,  
\label{BS4}
\end{equation}
and
\begin{eqnarray}
	& & [\tilde\lambda_{AA}(k,s) / \tilde\lambda_{AA} \, N(k,s)] =
	1 + [\tilde\lambda_{BB} (1 - \tilde\lambda_{AB} 
	\tilde\lambda_{BA} / \tilde\lambda_{AA} \tilde\lambda_{BB})]
						\, I_B(k,s) \ , 
\label{BS5} \\
	& & [\tilde\lambda_{BB}(k,s) / \tilde\lambda_{BB} \, N(k,s)] = 
	1 + [\tilde\lambda_{AA} (1 - \tilde\lambda_{AB} 
	\tilde\lambda_{BA} / \tilde\lambda_{AA} \tilde\lambda_{BB})] 
						\, I_A(k,s) \ .  
\label{BS6}  
\end{eqnarray}
At the normalisation point one has 
$I_i(2\mu,0) = C_d \mu^{-\epsilon} / \epsilon$, and after some tedious
but straightforward algebra eqs.~(\ref{BS4})--(\ref{BS6}) yield again 
the beta functions (\ref{betaa})--(\ref{betbb}).

We can now examine the above eqs.~(\ref{betaa})--(\ref{betbb}), which
we have just demonstrated to hold to {\it all} orders in perturbation
theory, for fixed point solutions $g_{ij}^*$ defined by 
$\beta_{ij}(\{ g_{ij}^* \}) = 0$. 
For $d > 2$ we find, as expected, merely the trivial Gaussian fixed
point where all $g_{ij}^* = 0$.
However, for $d<2$, the only {\it stable} fixed points are those 
describing {\it uncoupled} annihilation processes, i.e., 
\begin{equation}
        g_{AB}^* = g_{BA}^* = 0 \ , \qquad 
        g_{AA}^* = g_{BB}^* = \epsilon \ .
\label{scfp}
\end{equation}
Furthermore, there are also other solutions, for example the fixed line
\begin{eqnarray}
        & & 0 < c = g_{AB}^* \, g_{BA}^* \leq \epsilon^2 / 4 \quad
                {\rm fixed \ but \ arbitrary, \ and} \nonumber \\
        & & 2 g_{AA}^* = 
                \epsilon \pm \sqrt{\epsilon^2 - 4c} \ , \quad
        2 g_{BB}^* = \epsilon \mp \sqrt{\epsilon^2 - 4c}\ ;
\label{filine}
\end{eqnarray}
but these, like the Gaussian fixed point, turn out to be 
{\it unstable} for $d<2$.  

Hence the above annihilation / scattering model (\ref{scatt})
asymptotically becomes rather simple, and in fact lies in the same
universality class as single-species annihilation (with respect to
both the decay exponent {\it and} amplitude). 
Hence each species of particle decays according to eq.~(\ref{purann})
as $t \to \infty$ for $d < 2$. 
Physically this is a result of the `reentrancy' property of random
walks --- as soon as two particles are in close proximity, they will
rapidly annihilate, even in the presence of scattering processes. 
Therefore we conclude that, for this system, the presence of `real' as
well as `imaginary' noise has not introduced any novel behaviour.
We finally remark that the above results also apply in the extreme
asymmetric situation where, say, $\tilde\lambda_{BA} = 0$ but 
$\tilde\lambda_{AB} > 0$ originally, i.e., when there is spontaneous 
{\it unidirectional} transformation of pairs of $A$ particles into
pairs of $B$ particles, but not vice versa.
At least in this special case, our result that this pairwise
transmutation is irrelevant in the long-time limit, appears
nontrivial.
For example, in the related case of DP processes for $A$, $B$
particles with coinciding critical points, which are coupled via the
reaction $A \to B$, the usual DP critical exponent $\beta$ is replaced
by a much smaller density exponent as a consequence of an ensuing 
{\it multicritical} point \cite{coupDP}.

\section{The annihilation and fission process}

Our second reaction-diffusion system consists of the processes 
\begin{equation}
        A + A \to \emptyset \quad {\rm and} \quad 
        A + A \to (n+2) A \ , 
\label{fission}
\end{equation}
to which we assign the annihilation rate $\lambda$ and `fission' rate
$\sigma_n$. 
Note that these processes differ from the `Branching and Annihilating
Random Walks' \cite{GraKrau,TakTre,Tauber} mentioned earlier in that
offspring particles can only be produced upon collision of two $A$
particles.
The corresponding action derived from the master equation describing
the reactions (\ref{fission}) reads in terms of the {\it unshifted}
continuous fields ${\hat a}(x,t)$ and $a(x,t)$
\begin{equation}
        S = \int \! d^dx \int \! dt \left[ 
                {\hat a} (\partial_t - D \nabla^2) a
                - \lambda (1 - {\hat a}^2) a^2  
                + \sigma_n (1 - {\hat a}^n) {\hat a}^2 a^2 \right]
\label{fsfth}
\end{equation}
(the terms depending on the homogeneous, uncorrelated initial density
distribution and on the projection state have been omitted here). 
Once again we point out that this theory is valid only for
unrestricted particle occupation numbers in the microscopic model. 
It is quite possible that altering the microscopic rules for site
occupancy (for example by allowing only $0$ or $1$ particles at a
site) may change some of our later conclusions \cite{Malte}. 
If we now proceed by performing the shift ${\hat a} = 1 + {\bar a}$,
the effective action becomes 
\begin{eqnarray}
        & & S = \int \! d^dx \int \! dt \Biggl[ 
                {\bar a} (\partial_t - D \nabla^2) a
                + \left( 2 \lambda - n \sigma_n \right) {\bar a} a^2 
                + \nonumber \\ & & \qquad \qquad \qquad
                + \left( \lambda - {n (n+3) \over 2} \sigma_n \right)
                        {\bar a}^2 a^2 - \sigma_n \sum_{l=3}^{n+2} 
                        {n+2 \choose l} {\bar a}^l a^2 \Biggr] \ .
\label{fsshf}
\end{eqnarray}
{\it If} all vertices ${\bar a}^l a^2$ for $l \geq 3$ are neglected,
this field theory becomes equivalent to a nonlinear Langevin equation
\begin{equation}
        \partial_t a(x,t) = D \, \nabla^2 a(x,t) 
           + \left( n \sigma_n - 2 \lambda \right) a(x,t)^2 + 
                   a(x,t) \, \eta(x,t) \ , 
\label{fislan}
\end{equation}
\begin{equation}
        \langle \eta(x,t) \rangle = 0 \ , \quad \langle 
                \eta(x,t) \eta(x',t') \rangle = 
                \left[ n(n+3) \sigma_n - 2 \lambda \right] \,
                        \delta^d(x-x') \delta(t-t') \ ,
\label{fsnois}
\end{equation}
which again describes competition between `real' noise (associated 
with $\sigma_n$) and `imaginary' noise (associated with $\lambda$).
Upon comparing with the model of Ref.~\cite{Grinstein},
eqs.~(\ref{GE}), (\ref{GEnoise}), we see that the annihilation /
fission process apparently corresponds to their parameters $r=0$, 
$u = 2\lambda-n\sigma_n$, and $\nu = n(n+3)\sigma_n/2-\lambda$.

Notice that it is potentially dangerous to perform the shift 
${\hat a} = 1 + {\bar a}$ and then to arbitrarily omit certain
nonlinearities \cite{Cardy,Tauber}, due to the discrete symmetry 
${\hat a} \to - {\hat a}$, $a \to - a$, under which the action
(\ref{fsfth}) is invariant (for $n$ even). 
This symmetry corresponds to local particle number conservation modulo
2, which is lost in the Langevin description based on (\ref{fsshf}). 
Furthermore, the neglected terms have the same scaling dimension as
those retained. 
We therefore proceed with the analysis of the {\it unshifted} theory
(\ref{fsfth}). 
We find that both the annihilation and fission rate have identical
scaling dimension $\sim \mu^{2-d}$ and thus the upper critical
dimension is again expected to be $d_c = 2$. 
For $d > 2$, a description given by the mean-field equation
(i.e. eq.~(\ref{fislan}) without noise) should become qualitatively
correct. 
For $n \sigma_n < 2 \lambda$ this leads asymptotically to a density
decay ${\bar n}(t) \sim t^{-1}$ (with reduced annihilation rate 
$\lambda_R = \lambda - n \sigma_n/2)$; for $n \sigma_n > 2 \lambda$,
on the other hand, the density grows rapidly and diverges at 
$t_c = 1 / (n \sigma - 2 \lambda) n_0$, where $n_0$ is the initial
density. 
Thus, there is no stationary state in the active phase.

We now consider a one-loop analysis of the action (\ref{fsfth}) for 
$d \leq 2$. 
To this order all the couplings associated with the interaction
vertices in (\ref{fsfth}) are renormalised. 
For example the coupling $\sigma_n$ is renormalised
by the diagrams shown in Fig.~2a. 
However, in addition to this, other processes are also generated at
this order: $2 A \to n A$ (by a combination of fission and
annihilation, see Fig.~2b) and $2 A \to 2(n+1) A$ (by two successive
fission reactions, see Fig.~2c).
\begin{figure}
\begin{center}
\leavevmode
\vbox{
\epsfxsize=4.5in
\epsffile{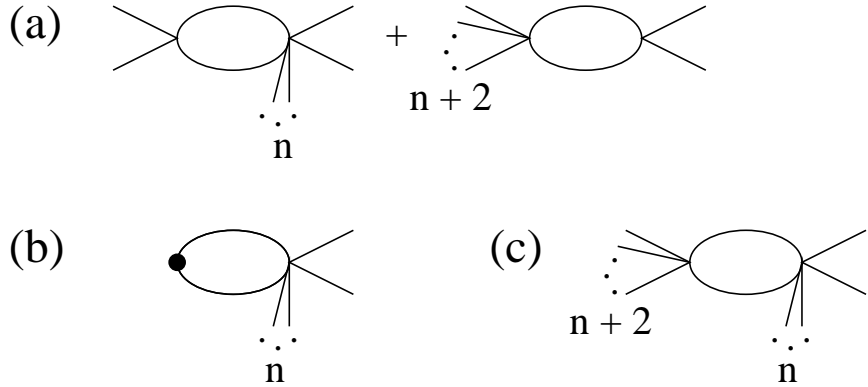}}
\end{center}
\caption{One-loop diagrams for (a) the renormalisation of the 
         $\sigma_n$ coupling, (b) the generation of the
	 process $2 A \to n A$ from a combination of $2 A \to (n+2) A$
         and $2 A \to \emptyset$, and (c) the generation of the
	 process $2 A \to 2(n+1) A$ by two successive fission
	 reactions.} 
\end{figure}
Furthermore, the mechanisms producing these processes (with
increasingly large $n$) are not simply restricted to the one-loop
level --- new particle creation vertices are effectively generated at
each successive order in perturbation theory. 
Therefore the number of (relevant) higher-order couplings, each of
which requires renormalisation, increases without bound as higher and
higher orders of perturbation theory are considered. 
Hence we must conclude that the field theory is non-renormalisable:
an infinite number of renormalisations would be needed to render the
theory free from divergences. 

Nevertheless some further progress is possible by considering the
original master equation for the annihilation-fission process
\cite{Gunter}.
Omitting the diffusive terms, we have
$$
{\partial P(\{m_i\};t)\over\partial t}=
        \lambda^{\rm lat}
        \sum_i[(m_i+2)
 	(m_i+1)P(\{\ldots,m_i+2,\ldots\};t)-m_i(m_i-1)P(\{m_i\};t)]
$$
\begin{equation}
	+\sigma_n^{\rm lat}\sum_i[(m_i-n)
	(m_i-n-1)P(\{\ldots,m_i-n,\ldots\};t)-m_i(m_i-1)P(\{m_i\};t)],
\label{masteq}
\end{equation}
where $P(\{m_i\};t)$ is the configuration probability for finding
occupation numbers $\{m_i\}$ at time $t$, and where 
$\lambda^{\rm lat}$ and $\sigma_n^{\rm lat}$ are the lattice
annihilation and fission rates, respectively. 
Using the relation $\overline{m(t)}=\sum_{\{m_i\}} m_i P(\{m_i\};t)$,
eq.~(\ref{masteq}) implies that
\begin{equation}
	{d\overline{m(t)} \over dt} =
	(n\sigma_n^{\rm lat}-2\lambda^{\rm lat}) \, \overline{m(m-1)}. 
\end{equation}
Since $m_i=0,1,2,\ldots$, we see that $\overline {m(m-1)}$ is
non-negative, and hence that $n\sigma_n^{\rm lat}=2\lambda^{\rm lat}$ 
marks the transition point between the active and inactive phases.
Note that actually at the transition ($n\sigma_n^{\rm lat}
=2\lambda^{\rm lat}$), the average density will remain
constant, whereas in the active phase it will diverge. 
These conclusions can be confirmed by studying the shifted action
(\ref{fsshf}). 
If we have the equality $2\lambda=n\sigma_n$ for the {\it bare}
field-theoretic parameters, then the bare cubic coupling 
vanishes. 
However, the structure of the higher-order vertices ensures
that a cubic coupling cannot then be regenerated {\it at any order} in
perturbation theory. 
Hence we can again conclude that $2\lambda = n\sigma_n$ is the
transition point between the active and inactive phases. 
Furthermore in the inactive phase, where the annihilation mechanism
dominates, and the successive generation of an infinite series of
fission processes is probably suppressed, we might expect the density
to decay as $t^{-d/2}$ (for $d<2$) due to the strong particle
{\it anti}correlations which emerge as a result of the annihilation
process in low dimensions.

However, the non-renormalisability of the field theory means that we
are unable to fully address the properties of either the active phase 
or the active / inactive transition. 
This failure may be associated with the fact that the active phase is
not in a stationary state (at least in mean-field theory, and with
unrestricted site occupancy $m_i = 0,1,\ldots,\infty$, the density in
this phase diverges in finite time). 
This behaviour, taken together with the massless nature of our field
theory (i.e., there is no term proportional to the field $a(x,t)$ in
eq.~(\ref{fislan})), implies that the transition from the absorbing to
the active phase cannot be in the DP universality class. 
Rather the transition is closer to being `first order', as suggested
by an exact evaluation of the correlation function at criticality
\cite{Gunter}. 
As field-theoretic methods clearly cannot shed any further light on
this problem, we hope that our analysis will stimulate further work
using, for example, exact one-dimensional methods.
In addition, numerical simulations presently in progress seem to
indicate that this annihilation / fission system may display
remarkably rich behaviour \cite{Malte}.

\section{Summary}

In this paper we have studied the effects of various types of noise in
diffusion limited reactions. 
In section 2 we emphasised that `naive' Langevin equations (with
positive noise correlators) fail to accurately describe systems
controlled by pair reaction processes, where the noise is in fact
`imaginary'.
Physically this failure is associated with the anticorrelation of 
particles in low dimensions. 
On the other hand such a naive approach does indeed work for the
Gribov process, where the noise turns out to be `real' (related to
particle clustering).

We then studied two diffusion-limited reaction systems with both real
and imaginary noise components: the annihilation / scattering
processes (\ref{scatt}) and the annihilation / fission processes
(\ref{fission}).
We have shown (to all orders in perturbation theory) that the first of
these belongs to the same universality class as the pure annihilation
model in dimensions $d \leq 2$, while for $d > 2$ the mean-field rate
equations apply.
However, the second system displays a transition between an active and
absorbing state, which is not accessible to perturbative analysis. 
In both cases, despite the competition between `real' and
`imaginary' noise, we have been unable to recover any of the
interesting behaviour discussed in Ref.~\cite{Grinstein}.
In fact, considering that the processes discussed here, along with
BARW, are amongst the {\it simplest} reactions leading to
both `real' and `imaginary' multiplicative noise, it is
rather unclear which {\it physical} system might be described
by the Langevin equation (\ref{GE}) with purely `real'
multiplicative noise (\ref{GEnoise}), and thus display the nontrivial
effects of Ref.~\cite{Grinstein}.

\vskip 1truecm

\noindent {\bf Acknowledgements.} 

\noindent
We would like to thank John Cardy, Geoff Grinstein, Hans-Karl Janssen
and Klaus Oerding for most interesting and inspiring discussions.
We are also grateful for correspondence with Malte Henkel, Fernando
Mendes and Gunter Sch\"utz who helped us identify an error in an
earlier version of this work, and to Ben Lee for useful comments. 
UCT acknowledges support by the European Commission through a TMR
Marie Curie Fellowship ERB FMBI-CT96-1189.



\end{document}